\documentclass[a4paper,fleqn,usenatbib]{mnras}
\usepackage{txfonts}

\usepackage[T1]{fontenc}
\usepackage{ae,aecompl}

\usepackage{threeparttable}
\usepackage{natbib}
\usepackage{adjustbox}
\usepackage{graphicx}
\bibpunct{(}{)}{;}{a}{}{,}
\usepackage{booktabs}
\usepackage[T1]{fontenc}
\usepackage{aecompl}

\title[]{Scl-1013644: a CEMP-s star in the Sculptor Dwarf Spheroidal Galaxy}
\author[Salgado et al.]{
C. Salgado,$^{1}$\thanks{Contact e-mail: \href{mailto:carolina.salgado@anu.edu.au}{carolina.salgado@anu.edu.au}}
G. S Da Costa,$^{1}$
D. Yong, $^{1}$
J. E Norris, $^{1}$
\\
$^{1}$Research School of Astronomy and Astrophysics, Australian National University, Canberra, ACT 2611, Australia\\
}

\date{Accepted XXX. Received YYY; in original form ZZZ}

\pubyear{2016}

\begin{document}
\label{firstpage}
\pagerange{\pageref{firstpage}--\pageref{lastpage}}
\maketitle

\begin{abstract}
Recent studies of the Milky Way and its satellites have paid special attention to the importance of carbon-enhanced metal-poor (CEMP) stars due to their involvement in Galactic formation history and their possible connection with the chemical elements originating in the first stellar generation. In an ongoing study of red giants in the Sculptor dwarf galaxy we have discovered a star with extremely strong CN and CH molecular bands. This star, Scl-1013644, has previously been identified by \citet{geisler2005sculptor} as a star with an enrichment in the heavy elements. Spectrum synthesis has been used to derive the carbon, nitrogen and barium abundances for Scl-1013644. Our findings are [C/Fe] = +0.8, [N/Fe] = --0.3 and [Ba/Fe] = +2.1 with the latter result consistent with the value found by \citet{geisler2005sculptor}. These results reveal Scl-1013644 as a CEMP-s star, the third such star discovered in this dwarf galaxy.
\end{abstract}


\section{Introduction}
\label{s_intro}

The study of metal poor stars provides clues about the early phases of star formation and chemical evolution in the universe. In order to understand the history of Galactic formation, many studies of metal-poor stars in the oldest structures, such as the Galactic halo and the ancient dwarf galaxies, have been made \citep[e.g.,][]{beers2005discovery, frebel2013metal, frebel2015near}.\\
Studies of the Galactic halo have shown that it has a population of carbon-rich stars, defined as [C/Fe] $ \geq $ 0.7 \citep{aoki2007carbon}, and that the frequency of these stars increases with decreasing metallicity \citep{beers2005discovery}. The definition of carbon-enhanced metal-poor (CEMP) stars encompasses different sub-groups: CEMP-s stars that show an excess of heavy elements produced by slow neutron capture (e.g. Ba); CEMP-r stars, which are dominated by rapid neutron capture elements (e.g. Eu); and CEMP-no stars which do not show any enhancement of neutron capture elements. Around 80$\%$ of the observed CEMP stars can be defined as CEMP-s stars making it the most common subclass of CEMP stars \citep{aoki2007carbon}. \\
The data suggest that the origin of CEMP-s stars is due to the mechanism of exchange of mass between a carbon enhanced AGB star and its binary companion, which is the star observed as the CEMP-s star today \citep{lucatello2005binary, cohen2006carbon, starkenburg2014binarity, izzard2009population}. The chemical abundances present in CEMP-s stars do not reflect the interstellar medium (ISM) from which they formed.\\
Many studies of carbon abundances in different environments have been performed. An example of these is \citet{kirby2015carbon}, who presented results for carbon abundances of red giants in several globular clusters and dwarf spheroidal galaxies. In their study 11 very carbon-rich stars were identified, of which 8 were previously known. The stars were found in three of the four dwarf galaxies studied but not in Sculptor. In this and the other dwarf galaxies \citet{kirby2015carbon} studied a large sample of red giant stars in order to understand the relation between the dSphs and the Galactic halo and their chemical enrichment. As regards the non-carbon enhanced stars ([C/Fe] < +1) in the dSphs, they compared the trend of [C/Fe] versus [Fe/H] with the trend for Galactic halo stars. They found that the `knee' in [C/Fe] occurs at a lower metallicity in the dSphs than in the Galactic halo.  \\
Additional interesting results for Sculptor have been obtained by \citet{skuladottir2015first} (hereafter AS15) who reported the first CEMP star in this dwarf galaxy, a star which does not show any overabundance of neutron capture elements. This finding adds to the previous CEMP-no stars found in Segue 1 \citep{frebel2014segue}, which has three CEMP-no stars, and Bo\"otes I which has one \citep{gilmore2013elemental}.
\\ Recently \citet{lardo2016carbon} (hereafter L16) presented carbon and  nitrogen abundance ratios and CH and CN index measurements for 94 RGB stars in Sculptor.  They reported that the [C/Fe] abundance decreases with increasing luminosity across the full metallicity range in the dSph, with the measurements of [C/Fe] and [N/Fe] in excellent agreement with theoretical model predictions \citep{stancliffe2009depletion}.  They also reported the discovery of two CEMP stars in Sculptor, both of which show an excess of barium consistent with s-process production. \\
Studies of carbon abundances have been also made in ultra-faint dwarf galaxies. Some remarkable results were found by \citet{norris2010extremely} and  \citet{frebel2014segue} in Segue 1. They found that the metal-poor red giants in this dwarf galaxy have the highest relative proportion of CEMP stars: of the 7 metal-poor red giants studied, four were found to be CEMP stars of which three stars are CEMP-no and one is CEMP-s.\\
In the current work we present further results for the Sculptor dwarf spheroidal galaxy, which is dominated by old (age $>10$ Gyr) metal-poor stars with no star formation having occurred for at least $\approx 6$ Gyr \citep{de2012star}. The galaxy is relatively faint, with $M_V\approx -11.2$, and has a distance from the Milky Way of $86\pm 5$ kpc \citep{pietrzynski2008araucaria}. 
As has been mentioned, many studies of carbon abundances have been made in Sculptor \citep[e.g.,][]{kirby2015carbon, frebel2010linking}. In general, the presence of carbon enhanced metal-poor stars is relatively infrequent in the classical (L $>10^5 L_\odot$) dSph companions  to the Milky Way  \citep{salvadori2015carbon}. \\
In this paper we report on the star \textit{Scl-1013644} (ID from \citet{kirby2010multi}), which has a metallicity [Fe/H] = --1.0, and which shows an substantial enhancement of carbon. This star also possesses a strong s-process element enhancement, which is compatible with the results from the high resolution study of \citet{geisler2005sculptor} (hereafter DG05). We have estimated the abundances of carbon, nitrogen and barium for \textit{Scl-1013644} from comparisons of the observed data with synthetic spectra.

\section{Observations and data reduction}
\label{s_obs}

Our sample of RGB members of the Sculptor dwarf spheroidal galaxy was selected from the work of \citet{walker2009stellar}, \citet{coleman2005absence}, \citet{battaglia2008analysis}. Our data are composed of 2 sets of observations that cover two distinct wavelength regions.  Spectra of the blue region, centred at 4300\AA, were collected with Gemini-South telescope on Cerro Pachon, Chile, using the Multi-Object Spectrograph (GMOS-S) \citep{hook2004gemini}. Six masks were observed with each mask observed twice at slightly different central wavelengths to compensate for the inter-chip gaps.
The red spectra, which cover 5800-6255\AA  \,\,approximately, were obtained with the Anglo-Australian Telescope (AAT) at Siding Spring Observatory using the 2dF multi-object fibre positioner and the AAOmega dual beam spectrograph \citep{saunders2004aaomega, sharp2006performance}. The red spectra are the combination of multiple 1800 second exposures obtained across 3 nights. Details of the observations are shown in Table \ref{Tabla_obs}. The blue spectra were reduced, extracted and wavelength calibrated with standard IRAF/Gemini software\footnote{\small{IRAF is distributed by the National Optical Astronomy Observatories, which are operated by the Association of Universities for Research in Astronomy, Inc.,  under cooperative agreement with the National Science Foundation.}}. These spectra permit the analysis of the G-band of CH ($\lambda\approx$ 4300\AA) and of the $\lambda\approx$ 3883\AA\/ and $\lambda\approx$ 4215\AA\/ bands of CN. The red spectra were reduced with the 2dF data reduction pipeline 2DFDR\footnote{\small{https://www.aao.gov.au/science/software/2dfdr}} and are useful for investigating s-process element abundances. Blue spectra are available for 45 Scl stars and red spectra are available for 161 Scl stars.\\

\begin{table}
\caption{Details of the observations.}
\resizebox{\linewidth}{!}{\begin{threeparttable}
\renewcommand{\TPTminimum}{\linewidth}
\centering 
\begin{adjustbox}{max width=8.5cm}
\begin{tabular}{l c c c r} 
\hline 
\hline
\small{} & \small{Blue (GMOS-S)} & \small{Red 	(AAOmega)} & \\ [0.5ex] 
\hline 

Grating				&	B 1200	&	2000 R 	&\\
Resolution ($\AA$)	&	2.3		&	0.8 		&\\
N\tnote{a}			&	45		&	161	&\\
Central Wavelength 	&	4400 \AA \,and 4350 \AA\tnote{b}	& 6050 \AA &\\

\hline 

\end{tabular}
\end{adjustbox}

\begin{tablenotes}
\item [a] Number of Sculptor stars with usable spectra. 
\item [b] Each mask was observed at two different central wavelengths.
\end{tablenotes}
\label{Tabla_obs} 

\
\end{threeparttable}}
\end{table}

\subsection{Line strengths}
\begin{figure}
\begin{center}
\includegraphics[width=8cm]{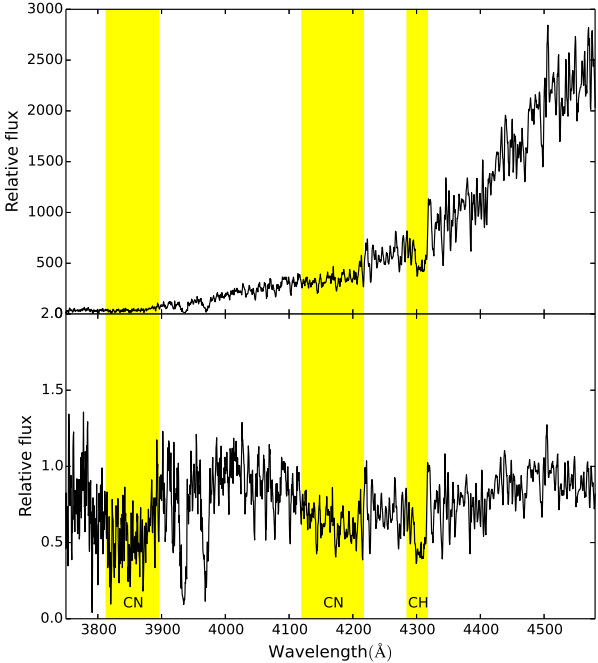}
\caption{Upper panel: Observed GMOS-S spectrum of Scl-1013644. Lower panel: Normalized spectrum of Scl-1013644. The locations of the CH, at $\lambda$ $\approx$ 4300\AA, and the CN, at $\lambda$ $\approx$ 3883\AA \/ and  $\lambda$ $\approx$ 4215\AA, bands are marked.}
\label{full_spectrum}
\end{center}
\end{figure}

When the reduced blue spectra were inspected we noticed immediately that the spectrum of star Scl-1013644 showed very strong CH and CN features.  The observed spectrum (not flux calibrated) and the normalized spectrum are showed in Figure \ref{full_spectrum}. The appearance of these spectra immediately allows the classification as a CH-star. The star has been included in previous studies under different IDs. It is star  \textit{10\_8\_2607} in \citet{coleman2005absence}, star \textit{982} in \citet{schweitzer1995absolute} and  \citet{geisler2005sculptor}, and \textit{Scl-1013644} in \citet{kirby2010multi}. These studies have unambiguously classified the star as a Sculptor member on the basis of radial velocity and/or proper motion.

\section{Syntheses and abundance analysis}

Prior to the abundance analysis it is necessary to normalize the continuum of the observed spectra (red and blue) of Scl-1013644. \\
The observed blue spectrum is the combination of two exposures taken at slightly different central wavelengths. 
As seen in the upper panel of Figure \ref{full_spectrum}, the extensive CH- and CN-bands make the normalization of the spectrum difficult. We fitted a low order polynomial to the (pseudo)-continuum points between $\approx$3900 and $\approx$4500 \AA. The resulting continuum normalized spectrum is shown in the lower panel of Figure \ref{full_spectrum}, and it has been used to compare with the synthetic spectrum calculations described below.\\
The observed red spectrum is a combination of a series of individual exposures obtained during the AAT observing run.  Because the wavelength coverage of the red spectrum lacks any strong CH- and CN- features, the continuum normalization was considerably easier.  As before a low order polynomial fit was used for the normalization.
\\
The abundance analysis has been performed using the local thermodynamic equilibrium (LTE) spectrum synthesis program MOOG \citep{sneden1973nitrogen, sobeck2011abundances} utilizing ATLAS9 model atmospheres \citep{castelli2003new}. We adopt solar abundances from \citet{asplund2009chemical}.\\
The stellar parameters used as input to generate the synthetic spectra are $T_{eff}$ = 3980 K, log\,$g$ = 0.5, [Fe/H] = --1.0, taken from DG05, and $\xi$ = 4.0 km s$^{-1}$. The stellar parameters have been also estimated by \citet{kirby2010multi} who lists: $T_{eff}$ = 4200 K, log\,$g$ = 0.41, and [Fe/H] = --0.96 dex.  These values are consistent with those of DG05. We have adopted the parameters estimated by DG05 in order to compare with more confidence our estimation of [Ba/Fe] and [O/Fe] with their values. The synthetic spectra of the CH and CN features were smoothed with a Gaussian kernel of 3.0 \AA \, FWHM. \\
Our results of the abundance analysis are presented in Table \ref{Tabla1}. Carbon, nitrogen, oxygen and barium abundances have been derived by comparison of the observed and theoretical spectra. The best fit is that which has the lowest residuals between the observed and synthetic spectra. Carbon was determined in the region of CH G-band between 4250 -- 4340 \AA. In this same region we investigated the effect of changing [O/Fe] on the derived [C/Fe]. Nitrogen was estimated using wavelengths between 4120 and 4290 \AA \, and the value for barium comes from the region 6138 -- 6144.5 \AA, which contains the $\lambda$6141.7\AA\/ line of BaII. Figure \ref{ch_cn} and Figure \ref{barium} show parts of these regions. We now discuss the details  of the abundance analysis.

\subsection{Carbon, Nitrogen and Oxygen abundances}
\label{CNO_abundances}

\begin{figure}
\begin{center}
\includegraphics[width=8.5cm]{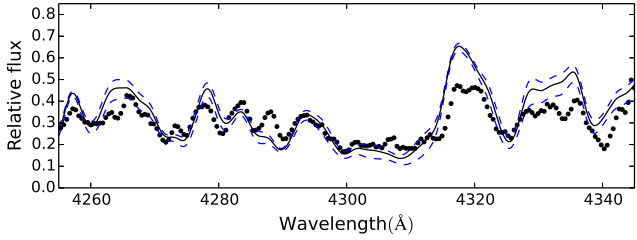}
\caption{Spectrum synthesis of CH. The observed spectrum is represented by dots. The solid line represents the best fit for abundances [O/Fe] = --0.3 (DG05) and [C/Fe] = --0.17. The dashed blue lines show carbon values $\pm$0.15 dex about the central value.} 
\label{ch_cn_geisler}
\end{center}
\end{figure}

\begin{figure}
\begin{center}
\includegraphics[width=8.5cm]{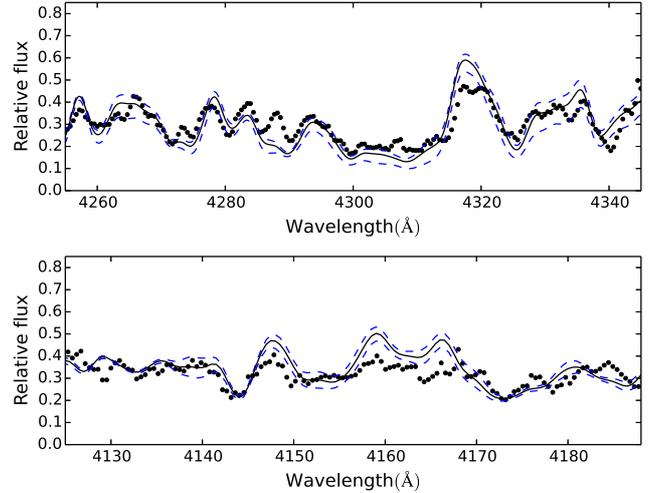}
\caption{Spectrum synthesis of CH (top panel) and CN (bottom panel) features for Scl-1013644. In both panels the observed spectrum is represented by dots. 
Upper panel: The solid line represents the best abundance fit with [C/Fe] = +0.8 and [O/Fe] = +0.7. The dashed blue lines show carbon values $\pm$0.15 dex about the central value.
Lower panel: The solid line represents the best abundance fit with [C/Fe] = +0.8, [O/Fe] = +0.7 and [N/Fe] = --0.3. The dashed blue lines show nitrogen values $\pm$0.2 dex from the best fit.}
\label{ch_cn}
\end{center}
\end{figure}

The abundance of oxygen (and nitrogen to a lesser extent) can affect the measurements of carbon because they form diatomic molecules with carbon and/or can contribute to the opacity of the stellar atmosphere. Unfortunately, it was not possible to measure these elements independently with our spectra. We performed the spectrum synthesis to estimate the carbon abundance, which was refined by exploring different assumed oxygen abundances.  In particular,
the sensitivity of the derived carbon abundance to the assumed oxygen abundance was tested by varying  the oxygen abundance from [O/Fe] = --0.3 to [O/Fe] = +1.3. The lower value corresponds to the value given by DG05 from the strength of the 6300\AA\,[OI] line. The upper value for [O/Fe] comes from the highest abundance of [O/Fe] found by \citet{kennedy2011fe} in their sample of 57 CEMP-s stars. Not surprisingly, we found that there was a substantial correlation between the [C/Fe] value derived from the synthetic spectrum fit and the assumed [O/Fe] value -- higher [O/Fe] values required a higher [C/Fe] for a satisfactory fit.\\
The procedure used to derive the final values for carbon and oxygen was firstly to make a rough estimation of the carbon abundance (assuming [O/Fe] = 0) based only on the similarities between the synthetic and observed spectra. The range of values that we considered reasonable was between [C/Fe] = +0.7 and [C/Fe] = +1.0. The next step was to refine the carbon abundance by introducing values of the oxygen abundance in the range defined above. We found that the best fit, which has the lowest residuals between the synthetic and observed spectra, is the pair [C/Fe] = +0.8 and [O/Fe] = +0.7.\\
Our attempts to use the value for oxygen given by DG05, [O/Fe] = --0.3, were unsuccessful due to poor agreement between the observed and synthetic spectra for this oxygen abundance. In Figure \ref{ch_cn_geisler} we show a synthetic spectrum for CH using [O/Fe] = --0.3 dex. In order to generate an acceptable fit it was necessary to change the value of carbon to [C/Fe] = --0.17 dex. However, such a low abundance ratio would appear to be inconsistent with the obvious strength of the CN and CH features in the observed spectrum shown in the panels of Figure \ref{full_spectrum}. \\
In order to investigate this difference further, we computed the strength of the oxygen line at $\approx$ 6300 \AA \, using our preferred oxygen abundance [O/Fe] = +0.7 and the oxygen value from DG05,  [O/Fe] = --0.3 dex.  In performing the calculation we used the same stellar parameters and resolution (R=22000) that DG05 adopted. Due to the strong presence of CN molecular features in the neighbourhood of the [OI] line, we also included in the synthesis our adopted values for carbon and nitrogen. A synthesis with  [O/Fe] = 0.74, [C/Fe] = 0.83 and [N/Fe] = --0.28 gives a equivalent width (EW) for the [OI] line of 72.1 m\AA. We repeated the synthesis calculation using the oxygen value given by DG05, [O/Fe] = --0.30, and used carbon and nitrogen abundances that allow an acceptable fit like that in Figure \ref{ch_cn_geisler}, namely [C/Fe] = --0.17 and [N/Fe] = --0.42. The EW of the [OI] line with this trio of abundances is  21.2 m\AA, which is roughly three times smaller than the EW using our abundance values. The EW of the 6300\AA\/ [OI] line listed by DG05 is 86 m\AA, which is completely consistent with our estimation with the large [O/Fe] value.
If we consider [O/Fe] = 0.74 then the location of this star in Figure 3 of DG05 becomes inconsistent with the trend for the other stars in the DG05 sample. However, we have to consider that the abundances of this star do not represent the original ones because it most likely has suffered a mass transfer from a companion. \\
To compare the quality of our fits with respect to the observational data we computed the root mean square (rms) of the difference between the observed and synthetic spectra for the fits shown in Figure \ref{ch_cn_geisler} ([C/Fe] = --0.17, [O/Fe] = --0.3) and in the upper panel of Figure \ref{ch_cn} ([C/Fe] = +0.83, [O/Fe] = +0.74). In the range 4315 -- 4345\AA,\, where we see larger differences, we found rms = 0.11 for Figure \ref{ch_cn_geisler} and rms = 0.07 for the upper panel of the Figure \ref{ch_cn}. 
If we consider the region 4250 -- 4345\AA, the results for the root mean squares are very similar, rms = 0.19 for Figure \ref{ch_cn_geisler} and rms = 0.18 for Figure \ref{ch_cn} upper panel. 
Quantitatively, the differences between the fits are not so strong in the range 4250 -- 4345\AA \,but in the region 4315 -- 4345\AA \, we see that the fit in the upper panel of Figure \ref{ch_cn} is significantly better than that in Figure \ref{ch_cn_geisler}.\\
The  CH($\lambda$\,4300) index for this star is CH($\lambda$\,4300) = 1.4, which was calculated using the same method as that employed by L16. With this value it is clear that Scl-1013644 does not lie with the  other relatively metal-rich giants in Figure 3 of L16.  Further, with our results for carbon and nitrogen, Scl-1013644 lies well away from the location of the Scl metal-rich giants in Figure 7 of L16.  Together these differences reinforce the unusual nature of the star and support our interpretation that it is a CH star.\\
Due to the difficulty in obtaining a correct normalization of the observed spectrum in the vicinity of the G-band, we have considered a range of carbon abundance of $\pm$0.15 dex from the central abundance of [C/Fe] = +0.8 to be the fitting error. \\
The nitrogen abundance was estimated from the region of the $\lambda$ $\approx$ 4216\AA\/ band of CN, more specifically in the region 4125 -- 4190 \AA. For the estimation of N we left fixed the values of carbon and oxygen. From the CN lines it is difficult to obtain an accurate value for the nitrogen abundance. Therefore, in order to get an estimate of the nitrogen abundance we show in the lower panel of Figure \ref{ch_cn} three nitrogen abundances that vary by $\pm$0.2 dex from the central value of [N/Fe] = --0.3 dex. The results are summarised in Table \ref{Tabla1}. 

\subsection{Barium abundance}
\label{barium_abundance}

\begin{figure}
\begin{center}
\includegraphics[width=8cm]{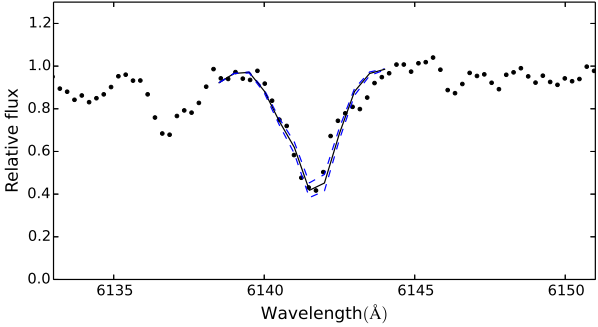}
\caption{Spectrum synthesis of the $\lambda$6141.7\AA\/ BaII line for Scl-1013644. The observed spectrum is represented by dots. The solid line represents the best fit which has [Ba/Fe] = +2.1. The dashed blue lines show barium values that differ by $\pm$0.15 from the best fit.}
\label{barium}
\end{center}
\end{figure}

CEMP-s stars are characterized by an excess of elements heavier than zinc and which are formed by the slow neutron capture process. The spectrum of Scl-1013644 around 6142 \AA \, shows a strong line of BaII (see Figure \ref{barium}) which may indicate a substantial s-process element enhancement. Such an enhancement is consistent with the definition of CEMP-s star. \\
To determine the Ba abundance we again used spectrum synthesis, smoothing the synthetic spectra to 1 \AA\ resolution to match the observations. The barium abundance was computed assuming solar isotopic ratios from \citet{mcwilliam1998barium} and hyperfine spitting (hfs). The neighbourhood of the BaII line is affected by lines of ZrI, SiI, and FeI among others but for the synthesis shown in Figure \ref{barium} we fit only the BaII line. For our synthesis we have assumed [Zr/Fe]= +0.7; however, if we use [Zr/Fe] = +1.2 from DG05 the fit does not noticeably improve.
The fit shown in Figure \ref{barium} is good but we also show in the Figure barium abundances that differ by $\pm$0.15 dex from the best fit. Our value of [Ba/Fe] = +2.1 is slightly higher than the value, [Ba/Fe] = +1.89, given in DG05 but considering the uncertainties these results are consistent.  Our derived abundance is again presented in Table \ref{Tabla1}.

\begin{table}
\caption{Derived abundances for Scl-1013644 ([Fe/H] = --1.0)}
\resizebox{\linewidth}{!}{\begin{threeparttable}
\renewcommand{\TPTminimum}{\linewidth}
\centering 
\begin{adjustbox}{max width=8.5cm}
\begin{tabular}{lcccr} 
\hline 
\hline
\small{Species} & \small{$log\epsilon\,_{Sun}$}  & \small{$log\epsilon\,_{star}$} & \small{[X/Fe]} & \\ [0.5ex] 
\small{(1)} & \small{(2)} & \small{(3)} & \small{(4)} & \\ [0.5ex] 

\hline

C (CH) 		& 8.43 & 8.26 	& + 0.83 	& \\
N (CN)		& 7.83 & 6.55   	& -- 0.28	 	& \\
O			& 8.69 & 8.43 	& + 0.74 	& \\
Ba	   		& 2.18 & 3.28	& + 2.10  	& \\

\hline 

\end{tabular}
\end{adjustbox}
\label{Tabla1} 
\end{threeparttable}}
\end{table}

\subsection{Error analysis}

The total error is the combination of the errors from the uncertainties in the stellar parameters, which we denote by $\sigma_{SP}$ and define in equation \ref{eq_sp}, together with additional sources of error, $\sigma_{fit}$, which include the effects of signal-to-noise and the fitting uncertainty.  The total uncertainty, $\sigma_{total}$, is then the combination, as given by equation \ref{eq_totalerror}.

\begin{equation}
\sigma_{SP} = \sqrt{\sigma_{T_{eff}}^2 + \sigma_{log g}^2 + \sigma_{\xi}^2 +
 \sigma_{[m/H]}^2}
\label{eq_sp}
\end{equation}

\begin{equation}
\sigma_{total} = \sqrt{\sigma_{SP}^2 + \sigma_{fit}^2 }
\label{eq_totalerror}
\end{equation}
\\
We have estimated $\sigma_{SP}$  by repeating the abundance analysis varying the atmospheric parameters by $T_{eff}$ = $\pm$200 K, log $ g$ = $\pm$0.2,  $[M/H]$ = $\pm$0.4 and   $\xi$ = $\pm$0.2 km s$^{-1}$. 
In the case of  the spectrum synthesis $\sigma_{fit}$ corresponds to the uncertainties of the fit in an rms sense.
It is important to remark that the strength of the CH molecular features is dependent on the oxygen abundance; therefore the C and O abundances are interdependent. They have been measured in the same spectrum synthesis and as a consequence of this, $\sigma_{fit}$ for C and O is the same. The $\sigma_{total}$ for carbon uses $\sigma_{fit}$ = $\pm$ 0.15 to which has been added a further $\sigma_{fit}$ = $\pm$ 0.15 from the assumed uncertainties in oxygen abundance.
As the determination of the nitrogen abundance from the CN features is subject to the adopted value of carbon, which is also dependent on the adopted oxygen abundance, we have considered this dependence as an additional source of error $\sigma_{add}$. Therefore, only for the $\sigma_{total}$ of nitrogen, we have added quadratically to the equation \ref{eq_totalerror} the value  of carbon $\sigma_{total}$ = 0.24  as an extra error.  \\
To summarize this section, we have considered the effects of errors in the stellar parameters, the error of the fits and additional sources (which just for nitrogen considers the dependence of C and O).  The values of these various uncertainties are listed in Table \ref{Tabla2}.

\begin{table}
\caption{Abundance errors from uncertainties in the atmospheric parameters and fits. }
\resizebox{\linewidth}{!}{\begin{threeparttable}
\renewcommand{\TPTminimum}{\linewidth}
\centering 
\begin{tabular}{lcccccccr} 
\hline
\hline
{species} & {$\Delta$ $T_{eff}$} &  {$\Delta$ log\,$g$	}  &  {$\Delta$[m/H]} &  {$\Delta$ $\xi$} &    {O} &  {$\sigma_{fit}$} &  {$\sigma_{Total}$} \\ [0.5ex] 
 {(1)} & {(2)} & {(3)} & {(4)} & {(5)} & {(6)} & {(7)} & {(8)} &\\ [0.5ex] 

\hline 

$\sigma$ C 		& 0.05 & 0.02 	& 0.10 	& 0.01 & 0.15 & 0.15  & 0.24  & \\
$\sigma$ N		& 0.05 & 0.02 	& 0.20 	& 0.01 & ---  & 0.20  & 0.37\tnote{a}  & \\
$\sigma$ O		& 0.07 & 0.02 	& 0.05 	& 0.01 & ---  & 0.15  & 0.17  & \\
$\sigma$ Ba   	& 0.01 & 0.07 	& 0.15 	& 0.09 & ---  & 0.15  & 0.24  & \\

\hline 

\end{tabular}
\begin{tablenotes}

\item [a] $\sigma_{Total}$ of nitrogen considers an extra error source  from carbon: $\sigma_{C}$= $\pm$ 0.24. 

\end{tablenotes}
\label{Tabla2} 
\end{threeparttable}}
\end{table}
\section{Discussion and Conclusion}
\label{s_concl}

In this work we present the discovery of Scl-1013644 as a previously unhighlighted CEMP-s star in the Sculptor dwarf galaxy. This star stood out in the sample of 45 stars observed because of the substantial strength of the CH and CN features in the star's spectrum. \\
An abundance analysis for carbon, nitrogen, oxygen and barium has been carried out for the star, via a comparison of the low resolution observed spectra from AAT/AAOmega and Gemini/GMOS-S with synthetic spectra models.\\
Scl-1013644 has [Fe/H] = --1.0 and our results indicate that it is enriched in carbon with [C/Fe] = +0.8 $\pm$ 0.3. The derived value for the nitrogen abundance is [N/Fe] = --0.3 $\pm$ 0.4 and, according to our estimate, the oxygen abundance is [O/Fe] = +0.7 $\pm$ 0.2. In addition, Scl-1013644 shows a strong  enrichment in barium with [Ba/Fe] = +2.1 $\pm$ 0.2. \\   
Scl-1013644 is located near the tip of the RGB in the colour-magnitude diagram. Therefore, due to the convective mixing that occurs in these stars, material processed through the CN-cycle has been incorporated into the surface layers of the star.  As a result, an enrichment in N and a depletion in C relative to the original abundances occurs at the surface.  It is therefore necessary to take this effect into account when considering the original surface C abundance before the onset of the mixing process.  Corrections to the observed C abundance as a function of evolutionary state on the RGB are discussed in \citet{placco2014carbon}.
For Scl-1013644, given the relatively high overall abundance, the evolutionary mixing correction to the observed C abundance is small, --0.02 dex, well within the uncertainty of the C abundance determination.  Nevertheless, with [C/Fe] = +0.8, Scl-1013644 does lie marginally above the limit adopted by \citet{aoki2007carbon}, confirming the star as a CEMP-star.\\
\citet{masseron2010holistic} present results of a statistical comparison  between [O/Fe] and [C/Fe] for carbon-enhanced metal-poor (CEMP) stars. Not surprisingly, this comparison shows that most of the stars studied, which cover a range in metallicities between [Fe/H] = --4.0 and [Fe/H] = 0.0,  are oxygen-enhanced relative to the solar abundance ratio. According to \citet{herwig2004evolution} the  overabundance of oxygen becomes larger with decreasing metallicity. Furthermore, that work states that the carbon and oxygen abundances are higher for lower initial masses. The overabundance of oxygen that we have found for Scl-1013644 is consistent with the \citet{masseron2010holistic} sample.\\
Many stars enhanced in carbon also show an enhancement of nitrogen. \citet{hansen2015elemental}, however, discuss a sample  of CEMP-s stars that have a [C/N] ratio greater than zero.  According to our findings, the [C/N] ratio for Scl-1013644 is larger than zero -- [C/N] = +1.1. The star then lies with other CEMP-s stars in Figure 4 of \citet{hansen2015elemental}. 
Our results of large [C/N] are also consistent with the model of AGB evolution described in \citet{herwig2004evolution} for an AGB progenitor of mass between 2--3 M$_\odot$.\\
Three CEMP stars are known in Sculptor from previous work (AS15, L16).  One is most likely a very metal poor CEMP-no star (AS15) while the other two are probably CEMP-s stars with [Ba/Fe] $>$ +1 (L16).  Our results for Scl-1013644 indicate that it is the third CEMP-s star in this dwarf galaxy, and it is considerably more metal-rich compared to the others.  Unlike the L16 stars, which have [C/Fe] = +1.5 and +1.4 and [C/N] = 0.6 and --0.2 for stars 20002 and 90085, respectively, Scl-1013644 has a smaller carbon enhancement but a higher [C/N] value and a higher overall metallicity.  CEMP-s stars are conventionally explained as the result of mass transfer to the current star when the original primary, of mass 2--3 M$_\odot$, evolves through the AGB phase.  That phase sees the dredge-up of substantial amounts of carbon and s-process elements such as Ba \citep[e.g.,][]{karakas2012slow}. We have no information concerning possible radial velocity variability for Scl-1013644, however, we see no reason to dispute the mass-transfer process as the likely origin for this and the other CEMP-s stars observed in Sculptor.  Once a complete sample of CEMP-s stars is established in Sculptor, it may be possible to use the frequency of occurrence of such stars to constrain the binary fraction in this dwarf galaxy \citep{starkenburg2014binarity}. Improved statistics on the occurrence of CEMP-no stars in this dwarf may also shed light on the likely different origin of such stars.
 

\section*{Acknowledgements}
\addcontentsline{toc}{section}{Acknowledgements}

Based on observations (Program GS-2012B-Q-5) obtained at the Gemini Observatory, which is operated by the Association of Universities for Research in Astronomy, Inc., under a cooperative agreement with the NSF on behalf of the Gemini partnership: the National Science Foundation (United States), the National Research Council (Canada), CONICYT (Chile), Ministerio de Ciencia, Tecnolog\'{i}a e Innovaci\'{o}n Productiva (Argentina), and Minist\'{e}rio da Ci\^{e}ncia, Tecnologia e Inova\c{c}\~{a}o (Brazil).
CS acknowledges support provided by CONICYT, Chile, through its scholarships program  CONICYT-BCH/Doctorado Extranjero 2013-72140033. This research has been supported in part by the Australian Research Council through Discovery Projects grants DP120101237, DP150103294 and FT140100554.
We thank Vinicius Placco for kindly computing the evolutionary correction to the carbon abundance.

\bibliographystyle{mnras}
\bibliography{cemp_mnras}

\bsp	
\label{lastpage}
\end{document}